\documentclass[longauth,traditabstract]{aa} 
\usepackage{graphicx}   
\usepackage{color}   
\newcommand{\micron}{\mbox{$\mu$m}}

\begin{document}   

\title{Cool and warm dust emission from M33 (HerM33es)\thanks{Herschel is an ESA space 
observatory with science instruments provided by European-led Principal 
Investigator consortia and with important participation from NASA.}
}

   \author{  
	E.M.\, Xilouris\inst{1} \and
	F.S.\,Tabatabaei\inst{2} \and
	M.\,Boquien\inst{3} \and
	C.\,Kramer\inst{4} \and
	C.\,Buchbender\inst{4}  \and
	F.\,Bertoldi\inst{5} \and
	S.\,Anderl\inst{5} \and
	J.\,Braine\inst{6} \and 
	S.\,Verley\inst{7} \and
	M.\,Rela\~{n}o\inst{7} \and
        G.\,Quintana-Lacaci\inst{4}  \and
        S.\,Akras\inst{8} \and
        R.\,Beck\inst{9} \and
        D.\,Calzetti\inst{10} \and 
        F.\,Combes\inst{11} \and
        M.\,Gonzalez\inst{4}  \and
        P.\,Gratier\inst{12} \and
        C.\,Henkel\inst{9,13} \and 
        F.\,Israel\inst{14} \and 
        B.\,Koribalski\inst{15} \and 
        S.\,Lord\inst{16} \and
        B.\,Mookerjea\inst{17} \and
        E.\,Rosolowsky\inst{18} \and
        G.\,Stacey\inst{19} \and
        R.P.J.\,Tilanus\inst{20,21} \and
        F.\,van der Tak\inst{22} \and
        P.\,van der Werf\inst{14,23}
}
   \institute{	
     Institute for Astronomy, Astrophysics, Space Applications \&
     Remote Sensing, National Observatory of Athens, 
     P. Penteli, 15236 Athens, Greece
     \and 
     Max-Planck-Institut f\"ur Astronomie, K$\ddot{\rm o}$nigstuhl 17,
     D-69117 Heidelberg, Germany	
     \and 
     Laboratoire d'Astrophysique de Marseille - LAM, 
     Universit\'{e} d'Aix-Marseille \& CNRS, UMR7326, 38 rue F. Joliot-Curie, 13388, 
     Marseille Cedex 13, France
     \and
     Instituto Radioastronomia Milimetrica (IRAM), 
     Av. Divina Pastora 7, Nucleo Central, E-18012 Granada, Spain
     \and 
     Argelander Institut fr Astronomie. Auf dem H\"ugel 71, 
     D-53121 Bonn, Germany
     \and 
     Laboratoire d'Astrophysique de Bordeaux, Universit\'{e} Bordeaux 1, 
     Observatoire de Bordeaux, OASU, UMR 5804, CNRS/INSU, B.P. 89, 
     Floirac F-33270
     \and 
     Departamento de F\'{i}sica Te\'{o}rica y del Cosmos, Universidad de Granada, 
     Campus Fuentenueva, Granada, Spain
     \and 
     Instituto de Astronom\'{i}a, UNAM, Campus Ensenada M\'{e}xico
     \and
     Max-Planck-Institut f\"ur Radioastronomie (MPIfR), 
     Auf dem H\"ugel 69, D-53121 Bonn, Germany
     \and
     Department of Astronomy, University of Massachusetts, Amherst,
     MA 01003, USA
     \and 
     Observatoire de Paris, LERMA, CNRS, 61 Av. de l'Observatoire, 
     75014 Paris, France
     \and
     Institut de Radioastronomie Millim\'etrique, 300 rue de la Piscine, 38406 Saint
     Martin d'H\`eres, France
     \and
     Astronomy Department, King Abdulaziz University, P.O. Box 80203, Jeddah, Saudi Arabia 
     \and 
     Leiden Observatory, Leiden University, PO Box 9513, 2300 RA Leiden, The Netherlands  
     \and 
     Australia Telescope National Facility, CSIRO, PO Box 76, Epping, 
     NSW 1710, Australia
     \and 
     Infrared Processing and Analysis Center, MS 100-22
     California Institute of Technology, Pasadena, CA 91125, USA
     \and 
     Department of Astronomy \& Astrophysics, 
     Tata Institute of Fundamental Research, 
     Homi Bhabha Road, Mumbai 400005, India
     \and 
     University of British Columbia Okanagan, 3333 University Way, 
     Kelowna, BC V1V 1V7, Canada
     \and 
     Department of Astronomy, Cornell University, Ithaca, NY 14853, USA
     \and 
     Joint Astronomy Centre, 660 North A'ohoku Place, University Park, 
     Hilo, HI 96720, USA
     \and
     Netherlands Organization for Scientific Research, Laan van Nieuw
     Oost-Indie 300, NL2509 AC The Hague, The Netherlands
     \and 
     SRON Netherlands Institute for Space Research, Landleven 12, 
     9747 AD Groningen, The Netherlands
     \and
     SUPA, Institute for Astronomy, University of Edinburgh,
     Royal Observatory, Blackford Hill, Edinburgh EH9 3HJ, UK
}	

   \offprints{E. M. Xilouris, \email{xilouris@noa.gr}}   
   \date{Received / Accepted }   
\abstract{
We study the far-infrared emission from the nearby spiral galaxy M33 in order to investigate 
the dust physical properties such as the temperature and the luminosity density across the galaxy.
Taking advantage of the unique wavelength coverage (100, 160, 250, 350 and 500~\micron) of the 
Herschel Space Observatory and complementing our dataset with Spitzer-IRAC 5.8 and 8~\micron~and
Spitzer-MIPS 24 and 70~\micron~data, we construct temperature and luminosity density maps by 
fitting two modified blackbodies of a fixed emissivity index of 1.5.
We find that the ``cool'' dust grains are heated at temperatures between 11 and 28 K with the lowest temperatures
found in the outskirts of the galaxy and the highest ones in the center and in the bright HII regions.
The infrared/submillimeter total luminosity (5 - 1000~\micron)
is estimated to be $1.9\times10^9$~$_{-4.4\times10^8}^{+4.0\times10^8}$ 
L$_{\odot}$.
59\% of the total luminosity of the galaxy is produced by the ``cool'' dust grains 
($\sim15$ K) while the rest 41\% is produced by ``warm'' dust grains ($\sim55$ K). 
The ratio of the cool-to-warm dust luminosity is close to unity (within the computed 
uncertainties), throughout
the galaxy, with the luminosity of the cool dust being slightly enhanced in the
center of the galaxy. 
Decomposing the emission of the dust into two components (one emitted by the diffuse disk of the 
galaxy and one emitted by the spiral arms) we find that the fraction of the emission in the disk
in the mid-infrared (24~\micron) is $21$\%, while it gradually rises up to $57$\% in the
submillimeter (500~\micron).
We find that the bulk of the luminosity comes from the spiral arm network that
produces 70\% of the total luminosity of the galaxy with the rest coming from the diffuse dust disk.
The ``cool'' dust inside the disk is heated at a narrow range of temperatures between 
18 and 15 K (going from the center to the outer parts of the galaxy).

}
\maketitle

\begin{figure*}[t]
  \centering
  \includegraphics[width=19cm,angle=0]{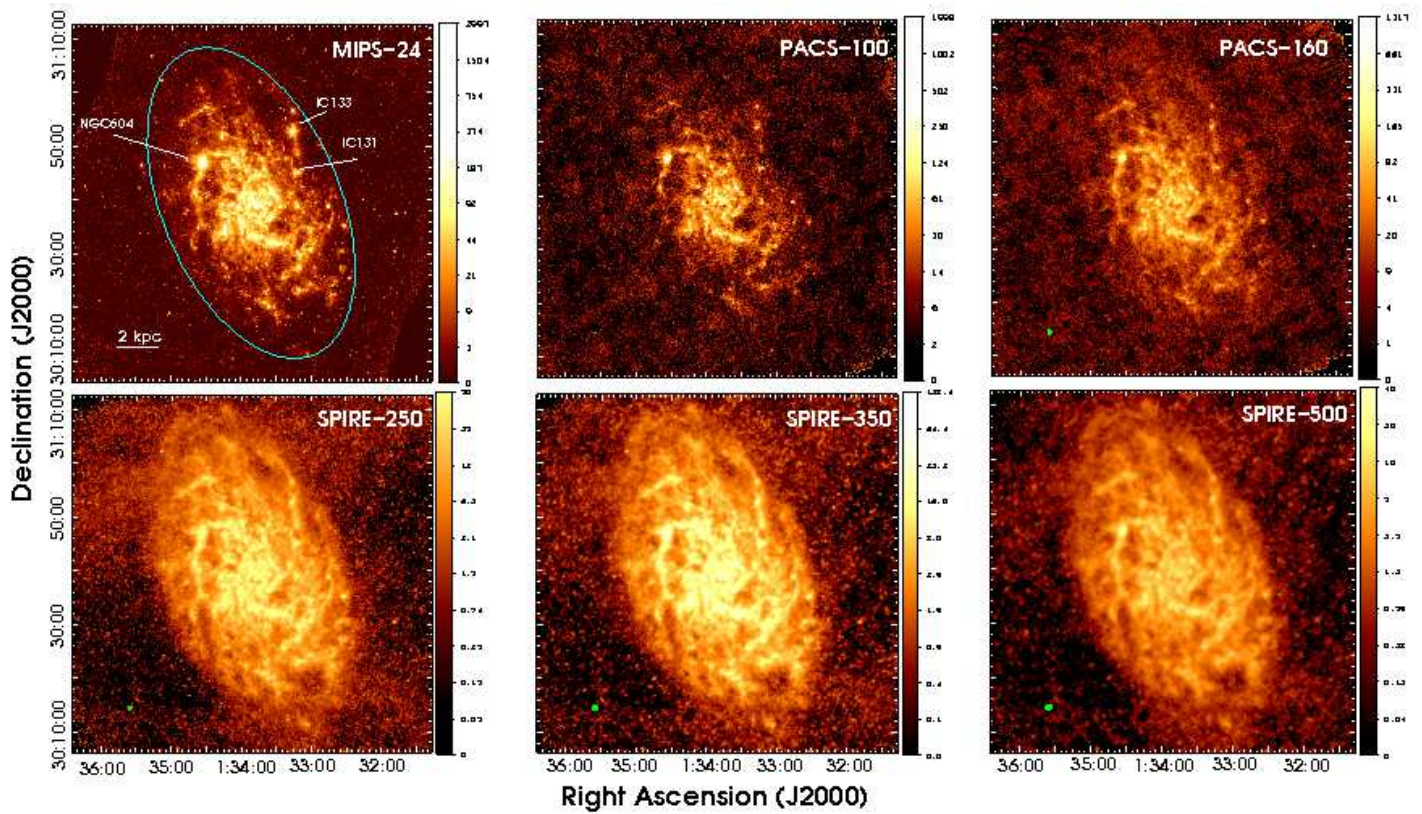}
  \caption{
Spitzer MIPS-24~\micron~(upper left panel), Herschel PACS-100 and 160~\micron~(upper middle and right panels, 
respectively) and SPIRE-250, 350 and 500~\micron~(lower left, middle and right panels, 
respectively). 
All maps are at their original resolution with the FWHM of the beams 
(6\arcsec, 
$6.7\arcsec \times 6.9\arcsec$, 
$10.7\arcsec \times 11.1\arcsec$, 
$18.3\arcsec \times 17.0\arcsec$, 
$24.7\arcsec \times 23.2\arcsec$, 
$37.0\arcsec \times 33.4\arcsec$
for 24, 100, 160, 250, 350, 500~\micron~respectively)
indicated as green circles at the bottom-left part in each panel.
In the MIPS-24~\micron~map, three of the brightest HII regions labeled for reference. 
The outer ellipse shows the B-band D$_{25}$ (semi-major axis length = 7.5 kpc; Paturel et al. 2003) extent of the galaxy.
The brightness scale is given in units
of MJy sr$^{-1}$.
}
\label{fig1}
\end{figure*}

\section{Introduction}
Studies of the dust content and of the properties of the 
interstellar medium (ISM) within 
nearby galaxies have been a major topic of research since the launch 
of the {\it Infrared Astronomical Satellite (IRAS)}. The IRAS measurements
at 60~\micron~ and 100~\micron, being sensitive in detecting the 
emission from the warm dust ($\sim$ 45 K) peaking at these wavelengths,
provided a unique database for studying this fraction of the
total ISM (e.g. Young et al., 1996, Devereux \& Young 1990). 
With the advent
of the {\it Infrared Space Observatory (ISO)} and the {\it Spitzer
Space Telescope}, measuring at wavelengths greater than
100~\micron, as well as the submillimeter (submm) observations
by the {\it James Clerk Maxwell Telescope (JCMT)}, it 
became evident that the bulk of 
the dust grains reside in a cooler (below 20 K) component 
making the most significant contribution to the total 
far-infrared (FIR) and submm emission of a galaxy. Although the existence 
of the warm and cool components is now well established,
spatial information on their distribution of their emission within galaxies
is still an issue. 
Galaxies showing prominent spiral structure in optical wavenegths show
strong dusty spiral arms as well as dust material diffusely distributed
throughout the galaxy revealing itself in the inter-arm regions 
(e.g., Haas et al., 1998; Gordon et al., 2006; Hippelein et al. 2003).  
A simplistic, nevertheless elegant, way to distinguish between 
the dust diffusely distributed throughout the galaxy and the dust
situated inside the spiral arms is to assume that the spiral
arm network is superimposed on an axisymetric dusty disk (e.g. 
Misiriotis et al. 2000; Meijerink et al. 2005). 
{\it Herschel Space Observatory} (Pilbratt et al.
2010) with its unique wavelength coverage, resolving power, and sensitivity 
is an excellent source of data for such studies.

The galaxies in the Local Group, because of their proximity,
are ideal systems for carrying out high spatial resolution studies 
of the interstellar medium (ISM). M33, an Sc spiral
galaxy at a distance of 840 kpc (Freedman et al. 1991), is an ideal candidate for the 
present analysis. Already mapped by ISO at 60, 100 and 170~\micron~
at a moderate resolution, M33 reveals a spiral structure
with a large number of distinct sources, as well as a diffuse
extended component (Hippelein et al., 2003). In the same study,
spectral energy distributions (SEDs) constructed for the 
central part of the galaxy as well as the interarm regions
and prominent HII regions, reveal typical temperatures of 
T$_{\rm w}\sim46$ K for the warm dust and T$_{\rm c}\sim17$ K for the 
cool dust component. 
Performing a multi-scale study of the infrared emission of M33
using Spitzer-MIPS data and a wavelet analysis technique, 
Tabatabaei et al. (2007a) concluded that most of the
24 and 70~\micron~ emission emerge from bright HII regions
and star-forming complexes, while the 160~\micron~ emission traces
both compact and diffuse emission throughout the galaxy.   
Using the intensity maps at 70 and 160~\micron~ Tabatabaei et al. (2007b)
constructed temperature maps showing variations of the temperature
between 19 and 28 K while similar conclusions were reached by
Verley at al. (2008) by constructing the dust temperature variations
as a function of radius from the center.

The recently launched Herschel Space Observatory  
offers the possibility to study M33 at FIR and submm
wavelengths in more detail than possible before. The PACS
(Poglitsch et al. 2010) and SPIRE (Griffin et al. 2010)
instruments combined allow to produce images over the
wavelength range between 70 and 500~\micron~ with unprecedented
sensitivity and superior resolution. 
In this paper we use
PACS (100~\micron~and 160~\micron) and
SPIRE (250~\micron, 350~\micron, 500~\micron) imaging,
taken as part of the Herschel M33 extended survey ({\tt HerM33es}; 
Kramer et al. 2010), an open time key project,
along with a two component modified blackbodies model in order to better 
understand the continuum emission of M33. 
With this approach we hope to shed some light on
the temperature and luminosity distributions of M33, 
being aware of the simplistic nature of the model that is used. A more
realistic analysis (within the framework of the {\tt HerM33es}
is presented elsewhere (Rosolowsky et al., in prep). 
In Section 2 we present the observations and the data reduction
techniques that we use, Section 3 gives a brief description of the
morphology of the galaxy at the Herschel wavelengths,
Section 4 describes modeling of the data, while the results of our
study are presented in Section 5 and discussed in Section 6.
We, finally, present our conclusions in Section 7.

\section{Observations and data reduction}
M33 was observed on 7 and 8 January 2010 with PACS and
SPIRE covering a field of $70\arcmin\times70\arcmin$. Observations were made
using the PACS and SPIRE parallel mode with a scanning 
speed of $20\arcsec/s$ providing simultaneous observations at
100~\micron~ and 160~\micron~ as well as at 250~\micron, 350~\micron, and 500~\micron.
The PACS maps presented in this paper were reduced using
the scanamorphos algorithm (Roussel 2012) 
discussed in detail in Boquien et al., (2011) and
in Rosolowsky et al., (in prep.). 
SPIRE maps were reduced using HIPE 7.0 (Ott 2010) and the SPIRE calibration
tree v. 7.0. SPIRE characteristics (point spread function, beam area,
bandpass transmission curves, correction factors for extended emission) were taken
from the SPIRE Observers' Manual (v2.4, 2011). A baseline algorithm 
(Bendo et al. 2010) was applied to every scan of the maps
in order to correct for offsets between the detector timelines
and remove residual baseline signals. Finally,
the maps were created using a ``naive'' mapping projection. 
For the errors in the PACS and SPIRE photometry we adopted a conservative
15\% calibration uncertainty for extended emission (Boquien et al., 2011; 
SPIRE Observers' Manual, v2.4, 2011).

In addition to the Herschel maps we also used the, publicly
available mid-infrared (MIR) maps, at 5.8, and 8~\micron~obtained with the IRAC
instrument, and the 24 and 70~\micron~obtained with the
MIPS instrument on-board the Spitzer Space Observatory. 
To account for the stellar pollution in the 5.8 and 8~\micron~fluxes
we used the IRAC-3.6~\micron~data as a reference point assuming that fluxes
at this waveband traces stellar emission only. This, however, is a
crude approximation since dust contamination, especially in the
bright HII regions, may be present in 3.6~\micron.
We then corrected 
the 5.8 and 8~\micron~maps by assuming a stellar light contamination 
dictated by the 3.6~\micron~measurements scaled with a Rayleigh-Jeans 
law. 

\begin{figure}[t]
  \centering
  \includegraphics[width=9.3cm,angle=0]{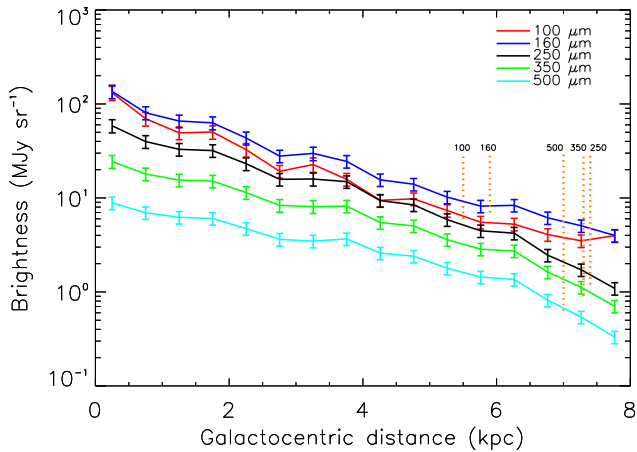}
  \caption{Azimuthally averaged radial profiles, projected on the major axis
of M33 for the 100, 160, 250, 350 and 500~\micron~emission (red, blue, black, green and
cyan colors respectively). The vertical dotted lines (in orange colour) indicate the
3-$\sigma$ flux level threshold for each wavelength.}
\label{fig1}
\end{figure}

The maps that were used in our analysis were all convolved 
to a common angular resolution
(the resolution of the 500~\micron~SPIRE image; approximately 40\arcsec)
by using the dedicated convolution kernels provided by Aniano et al. (2011)\footnote
{http://www.astro.princeton.edu/$\sim$ganiano/Kernels.html}
and projected to the same grid (of a 10\arcsec pixel) and center
position.

\section{Morphology}
Figure 1 shows the maps obtained at the five Herschel wave-
lengths (100, 160, 250, 350 and 500~\micron) as well as the 24~\micron~
map for comparison. 
In the 24~\micron~ map we have also marked the B-band D$_{25}$
(7.5 kpc; Paturel et al. 2003) extent of the galaxy
as well as the three of the brightest HII regions.
All maps are at their original resolution with the brightness
scale in MJy sr$^{-1}$.

\begin{figure}[t]
  \centering
  \includegraphics[width=9.3cm,angle=0]{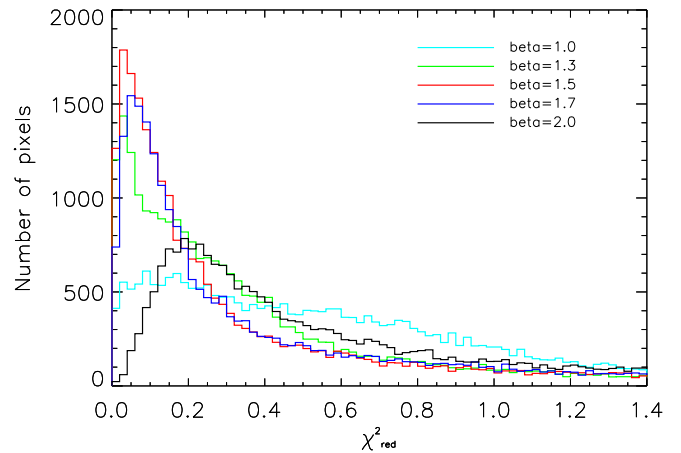}
  \caption{Histogram of the $\chi^2_{red}$ values for the three models with
$\beta$= 1, 1.3, 1.5, 1.7 and 2 (cyan, green, red, blue and black line respectively).}
\label{fig-maps}
\end{figure}

We use the Herschel maps to spatially 
model the SED across the galaxy (see Sect. 4).
All Herschel maps clearly show the spiral arm structure
and a large number of distinct sources. In addition to the
spiral structure, an extended underlying emission component
can be recognized (especially evident in the three SPIRE maps). This is
better illustrated by the azimuthally averaged radial profiles
presented in Fig. 2. The averaging in these profiles
is made within successive elliptical rings of 500 pc width
and a minor-to-major axis ratio of 0.54 accounting for an
inclination of 56 degrees (Regan \& Vogel 1994). 
The profiles are plotted all the way out to 8 kpc with the
orange vertical lines indicating the 3-$\sigma$ threshold in each band.
To these profiles we have fitted exponential functions
(minimizing the $\chi^2$) in order to derive the characteristic
scalelength of the dust emission at each wavelength.
We have done that for the range of radii with fluxes 
above the 3-$\sigma$ threshold.
We find scalelengths of 1.84 $\pm$ 0.12 kpc, 2.08 $\pm$ 0.09 kpc,
2.14 $\pm$ 0.08 kpc, 2.50 $\pm$ 0.12 kpc and 3.04 $\pm$ 0.15 kpc for the
100~\micron, 160~\micron, 250~\micron, 350~\micron~and 500~\micron, respectively.
The value for the scalelength that we derive for the 160~\micron~emission
using the Herschel-PACS data is in excellent agreement with the value
obtained using the Spitzer-MIPS data (1.99$\pm$0.02 kpc; Verley et al. 2009).
The flatter dust distribution at longer
wavelengths is indicative of a more extended distribution
which is related to emission dominated by ``cool'' ($\sim$ 15 K) dust grains
(see Sect. 6.1). Similar conclusions have been reached by others
using ISO (e.g., Alton et al. 1998; Davies et al. 199) and
Herschel (e.g., Bendo et al. 2012) observations.

The spiral arms in all Herschel bands can be traced all
the way to the center of the galaxy (to a 30 pc scale; which
is roughly the PACS 100~\micron~resolution). Two main spiral
arms, one to the north-east and one to the south-west, are
the brightest large-scale features of the galaxy. The
main spiral structure is also present at optical, near-infrared (NIR),
and radio wavelengths (Block et al. 2007; Tabatabaei et al.
2007b). Apart from these two open spiral arms our observations
show numerous, apparently independent, long-arm
spirals in the outer region of the galaxy, with a smaller pitch
angle. This appears like a flocculent spiral structure, while
in the NIR two arcs or wound spiral arms delineate another
coherent pattern. M33 is not the only galaxy showing a flocculent
spiral structure in young tracers, like gas and young stars,
superposed to a more grand-design spiral structure (e.g.
Block et al. 1996). These branched multiple spirals could
be random and independent wave packets triggered by local
gravitational instabilities in the gas and stars (Binney
\& Tremaine 1987) or they could result entirely from propagating
star formation and differential rotation (Gerola \&
Seiden 1978). In this case, gaseous instabilities trigger a
first star formation leading to supernova explosions and
shell-like compressions in the ISM. This triggers more star
formation leading to a chain reaction in which aggregates of
stars are created (see, e.g., Verley et al. 2010). The differential
rotation of the galaxy, then, stretches these aggregates
into spiral arcs. A characteristic picture of such violent processes
is also evident in M33 with a huge ($>$ 1 kpc) shell-like
feature which appears to break the south-west spiral arm
(evident in all Herschel maps of Fig. 1). A detailed analysis of the
morphology of the gas and dust content of M33 is presented
in Combes et al. (2012) using power-spectrum analysis techniques.

This picture is in excellent agreement with other tracers
of the ISM such as the atomic and the molecular hydrogen
(Braine et al. 2010; Gratier et al. 2010; Gratier et al. 2011). Many localized
infrared sources are well detected in the Herschel bands.
These sources are mainly HII regions situated in the spiral
arms showing enhanced star-formation activity. A detailed
analysis of the properties of these sources is presented in
Verley et al. (2010) as well as in Boquien et al. (2010).

\begin{figure}[t]
  \centering
  \includegraphics[width=10cm,angle=0]{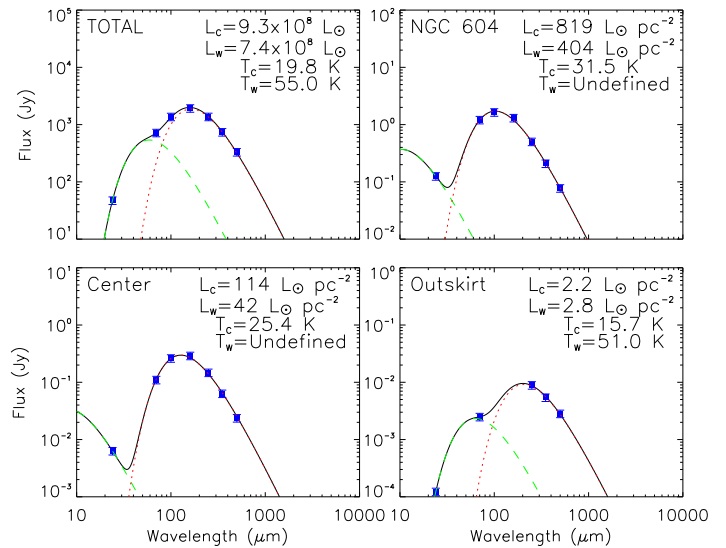}
  \caption{
SEDs, for the total galaxy (upper left panel) and for a typical HII region (upper
right - center of NGC 604), as well as for the center of the galaxy (lower left panel), and for
the ``diffuse'', outer part of the galaxy (lower right panel). The
emission was modeled with two modified blackbodies of $\beta$=1.5 and fitted to
the MIPS 24 and 70~\micron, PACS 100 and 160~\micron, and SPIRE
250, 350 and 500~\micron~observations. For the bright parts of the
galaxy (around the center and in bright HII regions) it is evident
that the specific model is not able to accurately describe the data giving
unphysically high temperatures for the ``warm'' dust component.
In the text we describe two different and more reliable methods to
derive the properties of the ``warm'' dust component
(see Sect. 4.1 for more details).
}
\label{fig-maps}
\end{figure}

\begin{figure}
  \centering
  \includegraphics[width=14cm,angle=0]{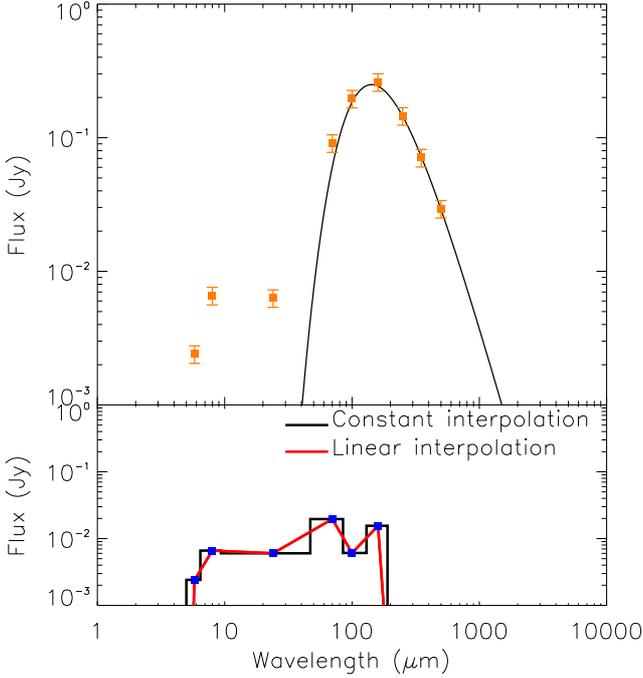}
  \caption{
Two of the techniques adopted to estimate the
``warm'' dust luminosity. After computing the ``cool'' dust
component (top panel) the residual MIR, FIR and submm fluxes
were calculated (blue points in the bottom panel).
These residual fluxes were then integrated using
(a) constant values within specified wavelength ranges according
to the filter passbands (black lines) and
(b) a linear interpolation between successive measurements
(red lines). The pixel used for this demonstration is
centered at 01:34:14.64, +30:34:31.57.
}
\label{fig-maps}
\end{figure}
\section{The model}
The unprecedented spatial resolution in the FIR and submm
wavelengths
obtained with Herschel allows us to study the dust grain
properties at a very small scale throughout
the galaxy. For this purpose the dust SED was modeled
assuming optically thin emission described by two modified blackbodies, one
accounting for the ``cool'' dust emission, originating from
the large grains being heated by the local and the diffuse
galactic radiation field, and one accounting for the ``warm''
dust emission, originating from the grains being heated in
dense and warm environments where star formation is
taking place.
According to this model the emission at a given frequency ($\nu$) is
described by:

\begin{eqnarray}
{\rm S}_{\nu} &=& \frac{1}{\lambda^{\beta}} [\rm N_{\rm c}B({\nu},T_{\rm c}) + 
N_{\rm w}B({\nu},T_{\rm w})]
\end{eqnarray}

\noindent
where $\rm B({\nu},T)$ is the Planck function, and T$_{\rm c}$, T$_{\rm w}$ are the
dust temperatures for the ``cool'' and the ``warm'' dust component
respectively, $\beta$ is the emissivity index while Nc and Nw are normalization constants
related to the dust column density in each of the ``cool''
and the ``warm'' components respectively.
To find the best SED for each pixel on the map,
our code minimizes the $\chi^2$ function 
($\chi^2 = \sum((S_{\nu, obs} - S_{\nu})/\Delta S_{\nu, obs})^2$)
using the Levenberg-Marquardt algorithm
(Bevington \& Robinson 1992).
The model was applied to the 24, 70, 100, 160, 250, 350 and 500~\micron~maps
regridded to a common resolution and pixel size (see Sect. 2) and for all pixels
with intensities larger than 3 times the rms noise level. 
The rms noise level for the (FWHM=40\arcsec and pixel size=10\arcsec) maps that
we used are 0.012, 0.31, 3.2, 3.4, 0.64, 0.50 and 0.32 mJy pix$^{-1}$
for the 24, 70, 100, 160, 250, 350 and 500~\micron~respectively. These
values were obtained by performing the statistics in regions well
outside the galaxy.
In order to expand our modeling to fainter parts of the galaxy (further than
the distances allowed by the 3-$\sigma$ criterium to all available observations)
we constrained the model by using the 24, 70, 250, 350 and 500~\micron~fluxes
only (i.e., dropping the low signal-to-noise 100 and 160~\micron~measurements
in the outer parts of the galaxy). These signal-to-noise considerations
allow for the 47\% of the galaxy's area to be modeled by using the full
dataset (7 wavelengths) while for the rest 53\% (mostly the outermost 
parts of the galaxy) the model is constrained by measurements in 5 wavelengths.  
We note here that, at this point of the analysis, we do not make use of
the 5.8 and 8~\micron~fluxes to constrain the  model.
These observations will be used later in order
to calculate the ``warm'' dust component of the SED (see Sect. 4.1).
To adjust for the filter bandpasses the SED was
convolved with the filter transmission before entering the $\chi^2$
minimization process. 
Based on the $\chi^2$ minimization, best
fitted temperatures and luminosity densities were obtained.

\begin{figure}
  \centering
  \includegraphics[width=9.3cm,angle=0]{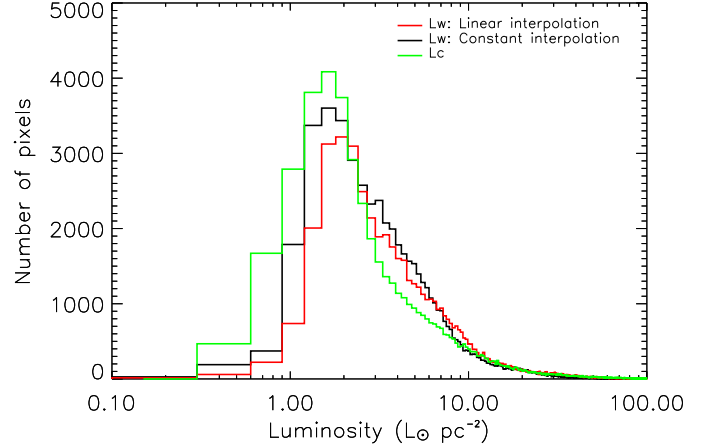}
  \caption{
The histogram of the luminosity densities of M33 which are
produced by the ``cool'' dust component (green) and the
``warm'' dust component (red and black).
The ``cool'' part of the luminosity was derived by fitting
the SED with a two modified blackbodies ($\beta=1.5$) model. The
``warm'' part of the luminosity is computed using
the residual fluxes of the SED, derived after
subtracting the ``cool'' dust component (see Fig. 5), and adopting
two approximations for integrating the residual SED by a
constant interpolation (black line) and a linear interpolation (red
line) between the successive fluxes (see the text for more details).
}
\label{fig-maps}
\end{figure}

In order to compute the uncertainty in the derived parameters we fitted the model
to fluxes obtained within the uncertainty range of the observed fluxes.
This was done by generating random deviates with a normal distribution (centered
in the observed flux; Press et al. 1986). For this purpose, 2000 such mock 
datasets were created which, after fitted by the SED model, resulted into
a set of 2000 values for each fitted parameter. From this distribution 
the 90\% confidence intervals were determined and lower and upper limits were
assigned on the best fitted value of each parameter.

The model dependence on the dust emissivity index $\beta$
was first explored by examining the change in $\chi^2$ for different
$\beta$ values. Fig. 3 shows the histogram of the reduced $\chi^2_{red}$
values (the $\chi^2$ divided by the number of observed parameters
minus the number of the fitted parameters minus one)
for $\beta$ = 1, 1.3, 1.5, 1.7 and 2. From this plot it is evident that the 
values close to $\beta$ = 1.5
account for more points with the minimum $\chi^2_{red}$ indicating
a better match between observations and model within the
error variance of the data. 
\begin{figure}[t]
  \centering
  \includegraphics[width=10.cm,angle=0]{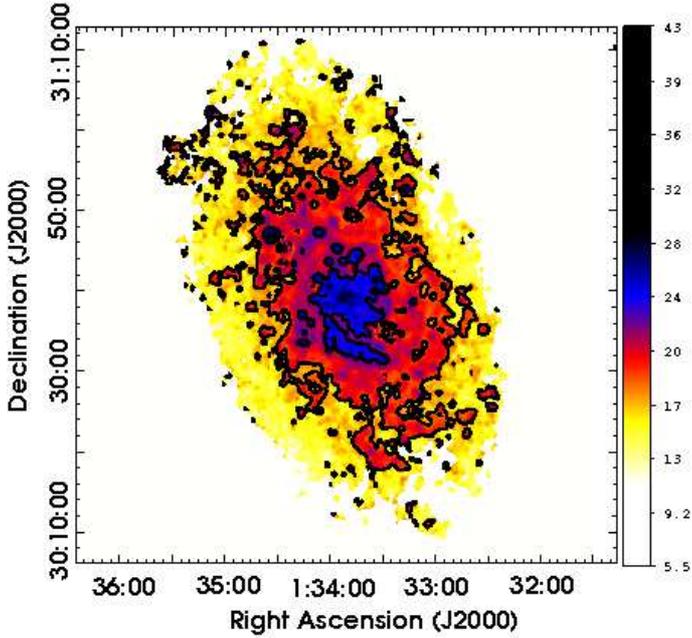}
  \caption{
The ``cool'' temperature (T$_{\rm c}$) distribution across the galaxy. The
two isothermal contours delineate regions with T$_{\rm c} < 18$ K
(yellow areas), 18 K $< T_{\rm c} < 23$ K (red/orange areas) and
T$_{\rm c} >$ 23 K (blue areas).
}
\label{fig-maps}
\end{figure}
The peak values of the histograms correspond to $\chi^2_{red}$ 
of $\sim$ 0.17, 0.02, 0.02, 0.06 and 0.19 for $\beta$=1, 1.3, 1.5, 1.7  and 2 
respectively with the $\beta=1.5$ distribution being the narrowest one.
This value of $\beta$ is in good agreement with the best fitting value found
by Kramer et al. (2010) in azimuthally averaged fluxes in elliptical
rings. It is also in good agreement with statistical studies 
of galaxies (e.g., Dunne \& Eales (2001), Yang \& Phillips (2007), 
Benford \& Staguhn (2008)).
A comprehensive analysis on the dust emission dependence with $\beta$
(within the framework of the {\tt HerM33es} Key Project)
is presented elsewhere (Tabatabaei et al. 2011).

\subsection{The ``warm'' dust component of the SED.}
The ``warm'' dust component is a highly unconstrained 
parameter of our model. This is because
the simple model that we use is not
capable of accurately describing the heating of the small 
dust grains emitting at short wavelengths. This is 
especially a very important caveat of
the model when fitting bright parts of the galaxy like the
central region and the cores of the bright HII regions. In
these regions, the largest part of the SED is adequately well
fitted by the ``cool'' component leaving only one flux measurement
(at 24~\micron) to constrain the ``warm'' part of the
SED. 
It is,
nevertheless, the unphysically large values of the ``warm'' dust
temperatures and luminosities as well as
the large uncertainties in these parameters
that make this model inadequate to
account for this part of the SED. Clearly,
a more realistic model, taking into account dust emission
at shorter, MIR, wavelengths is needed for a more complete
determination of the dust properties of the galaxy
(Rosolowski et al. in prep.).

\begin{figure}[t]
  \centering
  \includegraphics[width=9.3cm,angle=0]{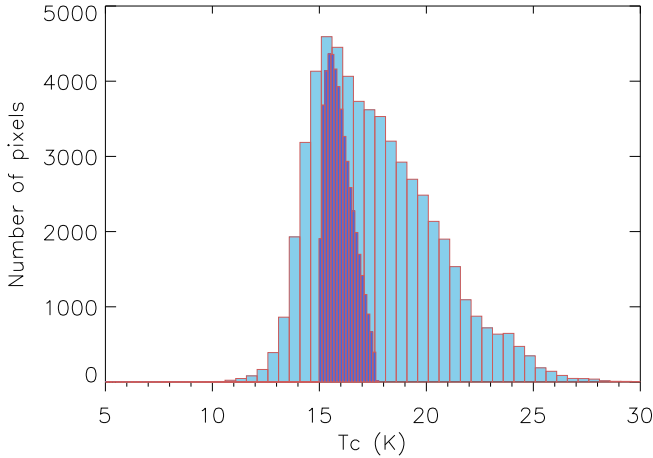}
  \caption{
The ``cool'' temperature (T$_{\rm c}$) histogram distribution.
The values derived for the whole galaxy are shown in the light blue
histogram while the
dark blue accounts only for the diffuse disk of the galaxy (see Sect. 6.2).
}
\label{fig-maps}
\end{figure}

Typical examples of single-pixel SEDs are shown in Fig. 4.
The upper right panel presents an SED near the 
peak of the NGC 604 HII region (01:34:34.09, +30:46:51.25),
the lower left panel shows an SED in the center of the
galaxy (01:33:45.21, +30:38:41.69), while a faint, diffuse, 
part in the outskirts of the galaxy (01:32:42.54, +30:30:50.58)
is presented in the lower right panel. The SED in the
last position shows a typical case where only five flux
measurements (PACS excluded; see Sect. 4) are used to constrain
the model. From these
SEDs the lack of data points shorter than 24~\micron~becomes
evident in the brightest parts of the galaxy (NGC 604 and center of the galaxy) with
the model failing in simulating the ``warm'' dust emission
giving unphysically large values for the temperature of this 
component.

In order to get an estimate of the energy output of M33
which is linked to the star forming activities that are taking
place inside the galaxy (the ``warm'' part of the SED) we
compute the luminosity of the residual fluxes (after having
subtracted the well defined ``cool'' component of the SED
as computed above; see Fig. 5). 
We do so, in a pixel by pixel basis, by
integrating the residual fluxes assuming a constant interpolation
between successive wavelengths (see an example in Fig. 5).
In particular, we assume
the 5.8, 8, 24, 70, 100, 160, 250, 350 and 500~\micron~residual
fluxes to be constant in the wavelength ranges of
5 - 6.44~\micron, 6.44 - 9.34~\micron, 9.34 - 47~\micron,
47 - 85~\micron, 85 - 130~\micron, 130 - 190~\micron, 190 - 300~\micron,
300 - 400~\micron~and 400 - 1000~\micron~respectively
(see the residual plot in Fig. 5). 
The wavelength ranges for the 5.8, 8, 250, 350 and 500~\micron~residual
fluxes were dictated by the actual width of the IRAC and SPIRE passbands
with the 500~\micron~range extended up to 1000~\micron. For the MIPS and
the PACS measurements we used the mid-point between successive
wavelengths as the wavelength limit.
With this method we compute a ``warm'' dust luminosity 
of $7.8\times10^8$~$_{-3.5\times10^8}^{+2.1\times10^8}$ L$_{\odot}$ of the entire galaxy. 
We note here that the well constrained ``cool'' dust component results in a luminosity production
of $1.1\times10^9$~$_{-2.0\times10^8}^{+3.9\times10^8}$ L$_{\odot}$.


In order to test the accuracy
of this method we also computed the ``warm'' dust
luminosity by integrating the residual fluxes but, this time,
assuming a linear interpolation for the residual fluxes at the
specific wavelengths (Fig. 5).
With this method we
compute a ``warm'' dust luminosity of $1.0\times10^9$~$_{-3.1\times10^8}^{+3.0\times10^8}$ L$_{\odot}$.
A more detailed comparison, on a pixel by pixel basis, between
the two methods for determining the ``warm'' dust luminosity, as 
described above, is given in Fig. 6 with the histograms of the luminosity
density (in units of L$_{\odot}$ pc$^{-2}$) presented for 
the two cases discussed above. 
For comparison, the histogram of the ``cool'' luminosity density 
is also plotted.
From these histograms, one can see that 
the distribution of the luminosity densities using the two
methods is very similar with the ``constant'' interpolation
approach giving a slightly broader distribution compared 
to the ``linear'' one.

A further check that we did in order to investigate the accuracy
of our method was to compute the luminosity
values for the two components by using
the total fluxes for the galaxy (integrated within an ellipse with a semi-major axis
length of 8 kpc).
With this approach both the ``cool'' and
the ``warm'' components could be calculated (see Kramer
et al. 2010) resulting in a ``cool'' dust luminosity of
$9.3\times10^8$~$_{-2.2\times10^8}^{+3.7\times10^8}$ L$_{\odot}$ and a ``warm'' dust luminosity 
of $7.4\times10^8$~$_{-5.4\times10^8}^{+2.6\times10^8}$ L$_{\odot}$ (see upper left
panel in Fig. 4).
   
\begin{figure}
  \centering
  \includegraphics[width=9.3cm,angle=0]{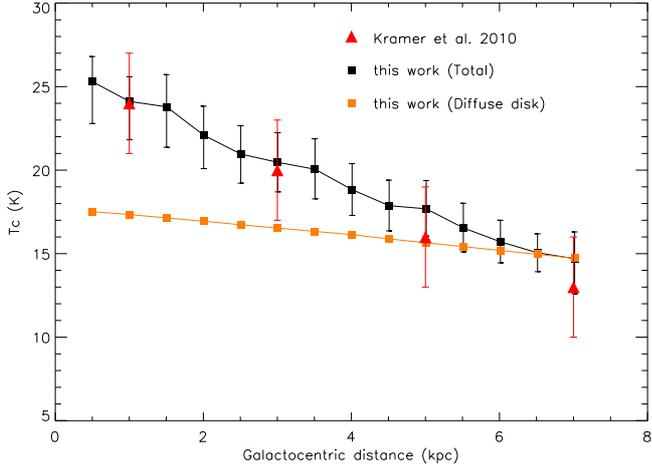}
  \caption{
Azimuthally averaged radial profile of the ``cool'' temperature
for the whole galaxy (black sqares) and for the disk only (orange
squares; see Sect. 6.2).
For comparison,
the ``cool'' dust temperatures for the dust that were calculated in Kramer et al.
(2010) for a two-component modified blackbodies model with $\beta=1.5$ and for the radial anulii:
$0<R<2$ kpc, $2<R<4$ kpc and $4<R<6$ kpc are also presented (red triangles).
}
\label{fig-maps}
\end{figure}

As a final test, we computed the total luminosity of M33 following
the recipe presented in Boquien et al. (2011). With this
approach the total IR luminosity was computed by fitting
the Draine \& Li model to the data (IRAC, MIPS, PACS,
SPIRE). As a product of this method, scaling relations were
produced between surface brightness measurements at specific
wavelengths and the total IR luminosity. For our purpose
we choose to use the 250~\micron~surface brightness map which
is the one with the best signal-to-noise. After calculating
the total IR luminosity and subtracting the already estimated
``cool'' dust luminosity we find a ``warm'' luminosity
of $8.4\times10^8 $ L$_{\odot}$. 

Given the large differences between the 
methods discussed above, the values for the ``warm'' dust luminosity are
very similar, within the computed uncertainties, giving us a high confidence that this part of the
luminosity, although estimated with very simple methods,
is an accurate first approximation. For the analysis to follow we make
use of the luminosity calculations derived with the ``constant'' interpolation
method since with this method the total ``warm'' luminosity is closer
to the values derived by both the Boquien et al. (2011) method and
the integrated fluxes method.
With this method the total luminosity of the galaxy (integrated in the 
5 - 1000~\micron~wavelength range) is 
($1.9\times10^9$~$_{-4.4\times10^8}^{+4.0\times10^8}$ L$_{\odot}$).

\section{Results}

\begin{figure}[t]
  \centering
  \includegraphics[width=10.cm,angle=0]{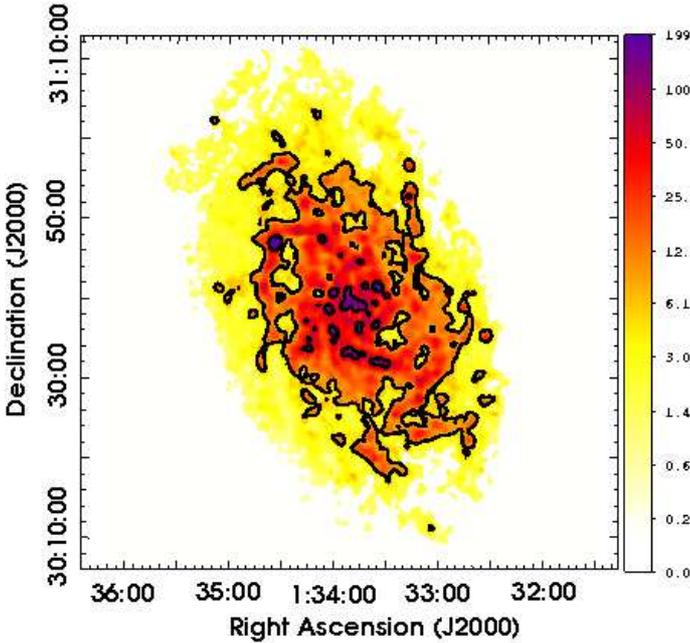}
  \caption{
The luminosity map of M 33 produced by the ``cool'' dust grains. Luminosities
less than 7 L$_{\odot} \rm pc^{-2}$  (yellow areas) are found in the
outer diffuse parts of the galaxy while
higher luminosities trace the spiral arms (orange/red areas). The highest luminosity
densities ($> 90$ L$_{\odot} \rm pc^{-2}$; innermost contours; blue areas) are produced
in the  bright HII regions and the center of the galaxy.
}
\label{fig-maps}
\end{figure}
\begin{figure}
  \centering
  \includegraphics[width=10.cm,angle=0]{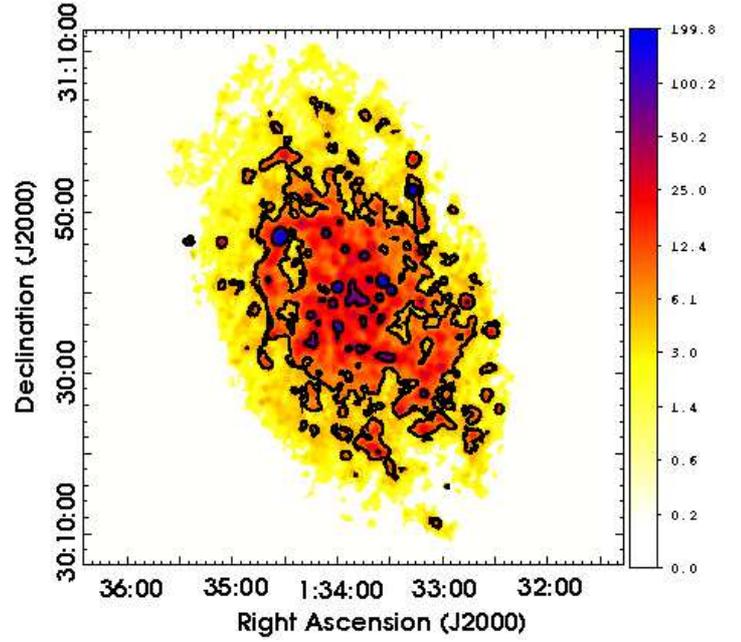}
  \caption{
The luminosity map of M 33 produced by the ``warm'' dust grains. Luminosities
less than 6 L$_{\odot} \rm pc^{-2}$ (yellow areas) are found in the outer diffuse 
parts of the galaxy while
higher luminosities trace the spiral arms (orange/red regions). The highest luminosity
densities ($> 40$ L$_{\odot} \rm pc^{-2}$; innermost contours; blue areas) are produced
in the  bright HII regions and the center of the galaxy.
The ``constant'' interpolation method was used to estimate
the ``warm'' dust luminosities.}
\label{fig-maps}
\end{figure}

Figure 7 shows the ``cool'' temperature pixel map created by modeling
the SED. This dust component
is heated at temperatures between 11 and 28 K
with $\sim$ 15 K being the most common temperature 
throughout the galaxy (see Fig. 8). 
In general, the ``cool'' temperature is symmetrically distributed
throughout M33 with typical temperatures of $\sim$ 25 K close to the center
of the galaxy, dropping to $\sim$ 15 K close to the outskirts.
The spiral structure is not clearly seen in the ``cool'' temperature 
distribution (Fig. 7). 
The effect of the dimming of the spiral arms with
respect to the large radial gradient of the galaxy is also present 
when looking at the 250/350 and 500/350
color maps (with the SPIRE colors being valuable indicators of
the ``cool'' dust temperature distribution; see Fig. 4
in Boquien et al. 2011 and Fig. 1 in Braine et al. 2010).
This is also evident when looking at the azimuthally
averaged profile of the temperature (black line in Fig. 9) showing a monotonic
decrease from the center to the outskirts with no obvious variations
from, e.g., the spiral structure. For comparison, the ``cool'' dust
temperature values calculated by Kramer et al. (2010) but for a range
of elliptical annuli are also plotted (red triangles). In general,
the values are within the errors for the temperature. 
The slightly higher temperatures that we calculate are due to
the small differences in the calibration of the Herschel 
data set that was used in Kramer et al. (2010).


Temperatures with T $<$ 18 K (yellow areas in Fig. 7) are found in the 
outer parts of the galaxy while larger temperatures trace the inner 
part of the galaxy. The highest
temperatures (T $>$ 23 K; innermost contour; blue areas in Fig. 7) are found in
individual HII regions and the center of the galaxy. This picture 
is also evident when examining
individual SEDs like the ones presented in Fig. 4 with
temperatures in the ranges discussed above for the HII regions 
(like, e.g., NGC 604), the center of the galaxy, and the diffuse 
outskirts regions of the galaxy. 
In these cases the best fit temperatures and the 90\%
confidence limits obtained with this model
are 31.5$_{-5.1}^{+1.6}$ K for the NGC 604 region, 25.4$_{-1.8}^{+1.0}$  K for
the central region of the galaxy, and 15.7$_{-3.0}^{+4.7}$ K in the 
outer, diffuse,  parts of the galaxy. These are
typical values of the temperature for the various regions
throughout the galaxy giving also an estimate of the accuracy
of the fitted parameters.
Given the simplicity of our SED model, and especially in the MIR
wavelengths (below $\sim$ 70~\micron), where significant contribution
to the emission from stochastically heated dust grains is expected, we
investigated the effects of omitting the 70~\micron~fluxes from the
analysis. In this way the ``cool'' dust component is constrained 
by observations at wavelengths greater than 100~\micron. We find that 
the ``cool'' dust temperature changes only slightly (variations of
less than 2\%) well within the estimated uncertainties. This gives us
enough confidence on the ``cool'' dust component calculation.

\begin{figure}[t]
  \centering
  \includegraphics[width=12.8cm,angle=0]{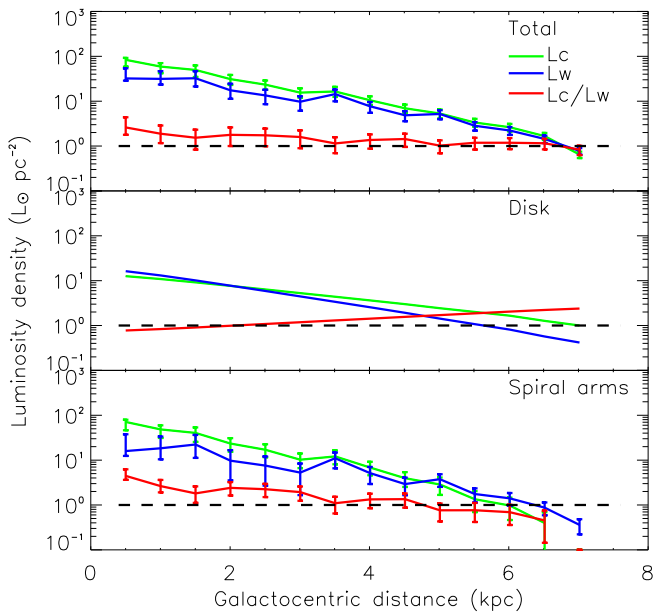}
  \caption{
The azimuthally averaged profiles of the ``cool'' and the ``warm''
dust luminosities (green and blue lines) in units of
L$_{\odot} \rm pc^{-2}$. The cool-to-warm luminosity ratio is
also plotted (red line). The luminosities are calculated for
the whole galaxy (upper panel) and for the disk of the galaxy only
(middle panel; see Sect. 6.2). The residuals between the two
profiles give the radial distribution of the luminosity produced
by the spiral arm network (bottom panel).
}
\label{fig-maps}
\end{figure}

For the ``warm'' dust component we do not present a temperature map since, 
as already stressed before, the temperature of this component, 
as derived from a two modified blackbodies fit, can be highly uncertain 
in many regions throughout the galaxy (especially the bright ones;
see the upper right and lower left panels in Fig. 4). It is 
worth mentioning though that, when considering the galaxy as a whole,
the ``warm'' dust temperature can be calculated and is 55.0$_{-3.7}^{+2.5}$ K
(in accordance to previous studies, e.g., Kramer et al. 2010) while for a 
typical, low brightness region (like the one in the lower right panel of
Fig. 4) the temperature is 51.0$_{-2.3}^{+2.9}$ K. 

The ``cool'' dust luminosity (Fig. 6 and 10) varies between
$\sim$ 0.3 L$_{\odot} \rm pc^{-2}$ and $\sim$ 800 L$_{\odot} \rm pc^{-2}$
with a value of $\sim1.7$ L$_{\odot} \rm pc^{-2}$ accounting for 
most parts across the galaxy.
Luminosities less than 7 L$_{\odot} \rm pc^{-2}$ (yellow areas in 
Fig. 10) are produced in the diffuse parts on the outermost regions of
the galaxy while higher luminosities (orange/red areas in Fig. 10) are emitted from the spiral arms. The 
highest luminosity densities (L $> 90$ L$_{\odot} \rm pc^{-2}$;
innermost contours (blue areas) in Fig. 10) are produced by the HII regions and the
center of the galaxy. This is also indicated when looking at individual
SEDs inside the galaxy (Fig. 4) resulting in a ``cool'' dust luminosity of
$819_{-415}^{+84}$ L$_{\odot} \rm pc^{-2}$ for NGC 604,
$114_{-24}^{+12}$ L$_{\odot} \rm pc^{-2}$ for the center and 
$2.2_{-0.95}^{+2.4}$ L$_{\odot} \rm pc^{-2}$ for the outer parts of the galaxy.

The ``warm'' dust luminosity (Fig. 6 and 11) varies between
$\sim0.3$ L$_{\odot} \rm pc^{-2}$ and $\sim$ 400 L$_{\odot} \rm pc^{-2}$
with a value of $\sim1.6$ L$_{\odot} \rm pc^{-2}$ accounting for
most parts across the galaxy.
Luminosities less than 7 L$_{\odot} \rm pc^{-2}$ (yellow areas in
Fig. 11) are produced in the diffuse parts of
the galaxy while higher luminosities are emitted from the spiral arms
(orange/red areas in Fig. 11). The
highest luminosity densities (L $> 40$ L$_{\odot} \rm pc^{-2}$;
innermost contours (blue areas) in Fig. 11) are produced by the HII regions
and the center of the galaxy.
This map has been computed by assuming the ``constant'' interpolation method
for estimating the ``warm'' dust luminosity (see Sect. 4.1). 
As in the case of the ``cool'' dust luminosity (discussed above) individual SEDs indicate
luminosities, produced by the ``warm'' dust material, of 
$404_{-67}^{+380}$ L$_{\odot} \rm pc^{-2}$ for NGC 604,
$42_{-27}^{+18}$ L$_{\odot} \rm pc^{-2}$ for the center and
$2.8_{-1.97}^{+0.6}$ L$_{\odot} \rm pc^{-2}$ for the outer parts of the galaxy.

\section{Discussion}

\subsection{The ``warm'' and the ``cool'' dust distribution}
Comparison of the ``cool'' and the ``warm'' luminosity maps in Figs 10 and 11
indicates that the morphology is very similar for the two dust components
with the spiral structure clearly seen in both cases and with local
enhancements in the areas where the HII regions reside.
This similarity becomes more striking when looking
at the azimuthally averaged radial profiles (upper panel in Fig. 12) of the luminosity densities of the
two components (green and blue lines for the ``cool'' and the ``warm'' components) 
and their ratio (L$_{\rm c}$/L$_{\rm w}$; red line). 
From this profiles it is evident that, given the uncertainties of the
computed luminosities, the ``cool'' dust luminosity is the dominant source
of luminosity in the central parts of the galaxy (with the ``cool'' luminosity being
$\sim3$ times more than the ``warm'' luminosity) while the ratio gets rather flat and 
close to unity beyond a $\sim3$ kpc radius.
This narrow range of the ratio (close to unity) of the two
luminosities suggests that both dust components (the ``cool'' dust heated at
$\sim$ 15 K and the ``warm'' dust heated at $\sim$ 55 K) are able to 
produce about the same amounts of luminosities throughout the galaxy.

\begin{table}
\caption{``Warm'' and ``cool'' dust luminosities for the
total galaxy, for the disk of the galaxy and for the spiral arms
of the galaxy.}              
\label{table:1}      
\centering                                      
\begin{tabular}{l}          
\hline\hline                        
Total galaxy\\    
\hline                                   
L$_{\rm{total}}$=1.9 $\times10^9$L$_{\odot}$\\      
L$_{\rm{w}}$=7.8 $\times10^8$L$_{\odot}$ (41\% L$_{\rm{total}}$)\\
L$_{\rm{c}}$=1.1 $\times10^9$L$_{\odot}$ (59\% L$_{\rm{total}}$)\\
\hline
Disk\\
\hline
L$_{\rm{disk}}$=5.7 $\times10^8$L$_{\odot}$ (30\% L$_{\rm{total}}$) \\
L$_{\rm{disk}}^{\rm{w}}$=2.5 $\times10^8$L$_{\odot}$ (32\% L$_{\rm{w}}$, 44\% L$_{\rm{disk}}$) \\
L$_{\rm{disk}}^{\rm{c}}$=3.2 $\times10^8$L$_{\odot}$ (29\% L$_{\rm{c}}$, 56\% L$_{\rm{disk}}$) \\
\hline
Spiral Arms\\
\hline
L$_{\rm{spiral}}$=1.3 $\times10^9$L$_{\odot}$ (70\% L$_{\rm{total}}$) \\
L$_{\rm{spiral}}^{\rm{w}}$=5.3 $\times10^8$L$_{\odot}$ (68\% L$_{\rm{w}}$, 40\% L$_{\rm{spiral}}$) \\
L$_{\rm{spiral}}^{\rm{c}}$=7.8 $\times10^8$L$_{\odot}$ (71\% L$_{\rm{c}}$, 60\% L$_{\rm{spiral}}$) \\
\hline                                           
\end{tabular}
\end{table}


Approximating the azimuthally averaged radial profiles of the
two luminosities with exponential functions, we find that they decay with
scalelengths of 2.59$\pm$0.04 kpc for the ``cool'' dust and 1.77$\pm$0.02 kpc
for the ``warm'' dust. Comparing with the scalelengths that we have derived
for individual wavelengths (Sect. 3 and this section) we see that the
``warm'' dust luminosity is very similar to the values derived for the
MIR/FIR wavelengths while the ``cool'' dust luminosity has a scalelength 
similar to the 250-500~\micron~submm wavelengths. 
This directly implies that the ``cool'' dust grains (mostly contributing 
to the ``cool'' part of the luminosity of the galaxy) emit in the 
submm wavelengths while the ``warm'' dust grains emit in the MIR/FIR
wavelengths shaping up the ``warm'' part of the luminosity of the galaxy. 

\subsection{Decomposing M33 into a disk and a spiral arm network}
The dust distribution inside a galaxy can be quite inhomogeneous. When considering
the morphology in a large scale the dust distribution can be approximated by a 
superposition of a diffuse disk and a 
spiral arm network (see e.g. Xilouris et al., 1999, Misiriotis et al. 2000;
Meijerink et al. 2005).
As a first order approximation the diffuse disk can be quite accurately
described by a three-dimensional disk decreasing exponentially both in the radial and in the 
vertical directions:
\begin{eqnarray}
\eta(R,z) &=& \eta_0 \exp(-\frac{R}{h_d}-\frac{|z|}{z_d})
\end{eqnarray}
where $\eta_0$ is the dust volume density at the
nucleus of the galaxy and $h_d$ and $z_d$ are the scalelength
and scaleheight of the dust respectively (see Xilouris et al. 1999). 
We make use of this simple model
to delineate between the two main dust components (i.e. the
the dust that is diffusely distributed in a disk throughout the galaxy 
revealing itself in the inter-arm regions of the galaxy and
the dust that is forming the spiral arms). 

\begin{figure}
  \centering
  \includegraphics[width=9.cm,angle=0]{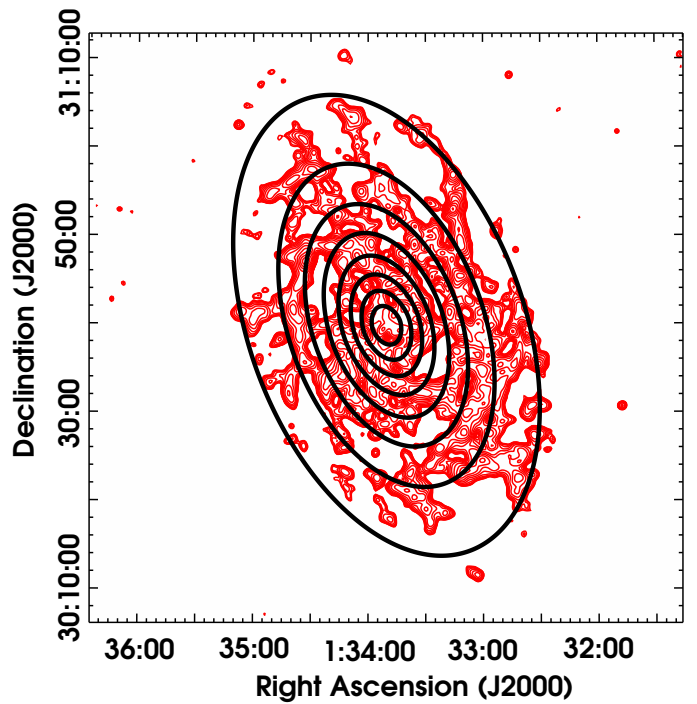}
  \caption{
The 250~\micron~image of M33 decomposed into a diffuse axisymmetric disk
(black contours) and its spiral structure network (red contours;
see the text for more details).
}
\label{fig-maps}
\end{figure}

\begin{figure}
  \centering
  \includegraphics[width=10cm,angle=0]{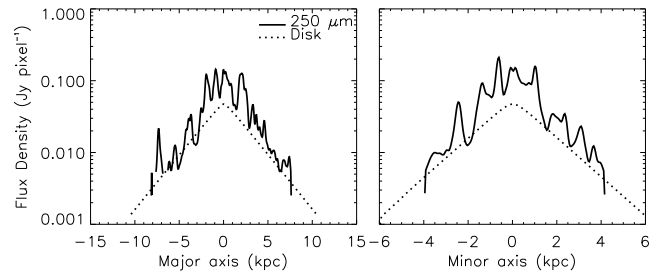}
  \caption{
The 250~\micron~profile of M33 along the major axis (left) and the
minor axis (right). In both cases the solid line corresponds to the
observed emission while the dotted line is the exponential disk
fitted to the data.
}
\label{fig-maps}
\end{figure}

Determining the scaleheight of the dust is not trivial since, unlike 
edge-on galaxies where the vertical extent of the dust can be measured, 
in less inclined galaxies, like M33, this parameter cannot be easily 
determined. We note however that for our study, where we aim at
calculating the relative fraction of the fluxes and luminosities
emitted by the diffuse and the spiral arm network, the choice of
the scaleheight is not a crucial issue. This is because the total
luminosity of the disk is independent of $z_d$ 
(a larger value of $z_d$ (thicker disk) would require a smaller value of $\eta_0$
(fainter disk) in order to fit the data which, at the end, results 
in the same luminosity; see Xilouris et al. 1997).
For the purposes of this study we use a scaleheight of $z_d=0.2$ kpc.
This is a typical scaleheight that describes the diffuse dust in a
galaxy derived from modeling and observations of edge-on galaxies
(Xilouris et al. 1999; Bianchi \& Xilouris 2011). In particular, 
for M33, this is close to the value derived by Combes et al. (2012)
using a power spectrum analysis.

For the scalelength we use the values determined by fitting the
azimuthally averaged profiles (see Sect. 3, Fig. 2). 
Since we also want to investigate the origin of the emission in
the MIR wavelengths we perform the analysis for the scalelength determination
(described in Sect. 3) in the MIPS-24 and 70~\micron. We find scalelength
values of $1.87\pm0.09$ and $1.90\pm0.05$ kpc respectively for the 24 and 70~\micron~emission.

With the scalelengths and scaleheights of the disk fixed the three dimensional 
model is then inclined by 56 degrees (the inclination angle of M33) and the
integrated, along the line of sight, intensities of the model are 
fitted to the galaxy maps in a pixel-by-pixel basis. We use only the fluxes
in the outermost regions of the galaxy and in the inter-arm regions
(the areas with minimum values in between the obvious spiral structure)
in order to constrain the emissivity at the galactic center ($\eta_0$).
We do this since we only want to fit the diffuse part of the galaxy and
avoid the spiral arms and the more complex structures that are superimposed on
the diffuse disk.
The fit was done by minimizing the $\chi^2$ values in a similar procedure
with what is described in Sect. 4.

An example of this analysis is presented in Fig. 13 showing the 250~\micron~map decomposition.
In this figure, the black contours show the axisymmetric exponential disk that
is fitted to the data (Eq. 2) while the red contours show the spiral arm network
determined after subtracting the map of the diffuse disk from the
observed image. 
In more detail this decomposition can be seen in Fig. 14 with the radial
profiles of the fitted disk to the 250~\micron~data (dotted lines) along 
with the observed major- and minor-axis profiles
(left and right panel respectively). From these profiles it is clear that 
the spiral arms are superimposed on the disk allowing for a clear separation 
of the two components.
 
Using the available Spitzer and Herschel maps of M33 we find that the flux 
that is emitted from the disk accounts for 
21, 29, 38, 42, 48, 52, and, 57\% of the
total flux for the 24, 70, 100, 160, 250, 350, and 500~\micron~respectively.
We see that in the MIR wavelengths it is the spiral structure which is the
dominant source of emission while, at around
$\sim$ 250~\micron~the emission comes equally from both the disk and the spiral 
arms. In contrast, the dominant source of the submm emission is
the disk of the galaxy.

Having calculated the emission that is produced by the diffuse dust
disk in all available bands, we can then calculate the 
temperature and the luminosity of this component separately.
We do so by applying the SED modeling that we did for the whole 
galaxy (Sect. 4) but now for this component only. 

For the ``cool'' temperature in the disk we find a very narrow distribution
(compared to the galaxy as a whole) with values ranging between 15 and 18 K.
This is shown in Fig. 8 with the dark blue distribution being that
of the diffuse disk. Given a typical 3 K uncertainty for the 
temperature distribution the disk temperatures occupy the lower end
of the temperature distribution for the whole galaxy (light blue) 
leaving the higher temperatures ($>18$ K)
accounting for the rest of the galaxy (spiral arms and HII regions).
The narrow range in temperatures is also seen in Fig. 9 with temperatures of
$\sim18$ K in the center of the galaxy dropping to $\sim15$ K at the edges.
A similar narrow range of temperatures for the diffuse dust in the
disk of the Milky Way has been derived  based on COBE observations.
Temperatures for the diffuse cirrus component ranging from 17 K to 21 K are
found using FIR color maps (e.g. Reach et al. 1995; Schlegel et al. 1998)
while a 15 K to 19 K range has been computed for the disk of
the Milky Way via radiative transfer modeling (Misiriotis et al. 2006).
Similar findings have been recently derived using Herschel observations 
of the galactic plane indicating that the warmest dust is located in the
spiral arms (Bernard et al. 2010). The narrow range of the temperatures that
we find in the disk of M33 is consistent with this picture, with the disk 
primarily hosting the cirrus dust component which is heated by the 
diffuse interstellar radiation field. On the other hand, the dust in the spiral arms 
is embedded in more dense environments, and is, mainly, heated by the
local UV radiation of the young stars, allowing for a wider range of
temperatures. We note however that a more detailed analysis, taking into 
account both the
exact radiation field that is heating the dust as well as the dust 
composition and density in various places throughout the 
galaxy is clearly needed for a 
proper characterizing the properties of the interstellar dust in
different environments.    

Concerning  the luminosity, we find that $5.7\times10^8$ L$_{\odot}$ (30\%
of the total luminosity) is produced by the dust in the  disk. This 
indicates that the dominant source of
the luminosity of the galaxy in the infrared/submm wavelengths
is the spiral arm network. 
Furthermore, looking at the ``warm'' and the ``cool'' components 
separately we compute that the dust that resides in the disk and is ``warm'' 
emits $2.5\times10^8$ L$_{\odot}$ which is 
32\% of the total ``warm'' dust luminosity while
the ``cool'' dust material produces $3.2\times 10^8$ L$_{\odot}$ which
is 29\% of the total ``cool'' dust luminosity. 
All the above results, as well as the values
derived for the spiral arms, are summarized in Table 1 

The above results indicate that when considering the luminosities of the 
``warm'' and the ``cool''
dust their origin is mainly the spiral arm network which
produces 71\% of the luminosity of the ``cool'' and 68\% of the luminosity of the ``warm''
dust of the galaxy. Furthermore, 41\% of the total luminosity
of the galaxy is emitted by ``warm'' grains while the rest 59\% is
emitted by ``cool'' dust grains. These percentages are about the same when
separating the disk from the spiral arms. We see, for example, that inside 
the disk 44\% of the luminosity is emitted by the ``warm'' dust compared 
to 56\% which is emitted by the ``cool'' dust
and, similarly for the spiral arms, 40\% comes from the ``warm'' and 60\% from the ``cool''
dust. 

It is interesting to further investigate the relative distribution of the
``warm'' and ``cool'' luminosities that reside inside the disk and the spiral arms 
as a function of galactocentric radius. The decomposition of the azimuthally
averaged radial profiles is shown in Fig. 12 with the disk profiles 
presented in the middle panel and the spiral arm profiles in the bottom panel.
The disk profiles (middle panel) are derived by fitting the two
modified blackbodies model to the smooth disk maps in each available wavelength.
These maps are derived by fitting the three-dimensional model (Eq. 2) 
to the observations as described earlier.
From these profiles it is evident that the ``cool'' dust in the disk of
the galaxy becomes the dominant component in the outer parts of the galaxy (beyond
$\sim3$ kpc) while the ``cool'' dust is the main source of luminosity production
within the inner $\sim3$ kpc inside the spiral arm network.

\section{Conclusions}
Through modeling of the SED of the local group galaxy M33 and exploiting the
unique wavelength coverage and sensitivity of the Herschel 100, 160, 250, 350
and 500~\micron~observations, as well as additional Spitzer-IRAC 5.8 and 8~\micron~
and Spitzer-MIPS 24 and 70~\micron~data, we derive the following conclusions on the 
dust emission of the galaxy:
\begin{itemize}
\item{The submm emission from M33 is distributed in a
flatter way (scalelength of 3.04 kpc at 500~\micron) compared to the far-infrared 
emission (scalelength of 1.84 kpc at 100~\micron). 
}
\item{On average, the emission by the dust can be adequately modeled by a superposition
of two modified blackbodies with $\beta=1.5$.}
\item{The temperature of the ``cool'' grains ranges between 11 and 28 K with 
a most common value of $\sim15$ K across the galaxy.}
\item{The total luminosity of the galaxy (integrated from 5 to 1000~\micron) is
$1.9\times10^9$~$_{-4.4\times10^8}^{+4.0\times10^8}$ L$_{\odot}$
with the luminosity of the ``cool'' dust component
accounting for 59\% of the total luminosity.}
\item{The scalelength of the ``cool'' (``warm'') luminosity distribution is very similar
to that of the submm (MIR/FIR) emission. This similarity directly reflects the 
important wavelengths that contribute the most in each of the two
luminosity components.}
\item{Decomposing the galaxy into two components (namely, a diffuse disk and a
spiral-arm network) we find that the emission
in the disk accounts for $\sim21$\% in the MIR wavelengths (24~\micron) while it 
gradually rises to
$\sim57$\% in the submm wavelengths (500~\micron).}
\item{The ``cool'' dust material in the disk of the galaxy, presumably accounting
for the cirrus dust component of the galaxy, is
heated at temperatures between $\sim$ 15 K at the edges of the galaxy to $\sim$ 18 K
at the center.}
\item{The bulk of the luminosity is emitted by the spiral arm network ($\sim$ 70\%) and
this percentage stays roughly the same when considering the ``warm'' and the ``cool''
dust luminosities separately.}
\item{About $\sim40$\% of the total luminosity throughout the galaxy (and separately 
inside the disk and the spiral arms) is emitted by the ``warm'' dust grains with the
rest $\sim60$\% emitted by the ``cool'' dust.}
\item{The ratio of the ``cool'' to ``warm'' dust luminosity is close to
unity throughout most of the the galaxy but slightly enhanced within the central $\sim$ 3
kpc of the galaxy.}
\item{The ``cool'' dust inside the disk becomes dominant in the outer parts of
the galaxy (beyond $\sim3$ kpc).}
\item{The ``cool'' dust inside the spiral arms become dominant in the inner
parts of the galaxy (within $\sim3$ kpc).}
\end{itemize}

\begin{acknowledgements} 
We are gratful to Marc Sauvage, George Bendo, Pierre Chanial, Michael Pohlen
and Richard Tuffs for their help in the data reduction. We also thank the anonymous
referee who provided useful comments on improving the paper.
\end{acknowledgements}   
   

\end{document}